\documentclass{article}
\usepackage{amssymb,amsfonts,amsthm,graphicx,psfig}
\usepackage{amssymb}
\usepackage{amsthm}
\usepackage{amsmath}
\def\pmb#1{\setbox0=\hbox{$#1$}%
  \kern-.025em\copy0\kern-\wd0
  \kern.05em\copy0\kern-\wd0
  \kern-.025em\raise.0433em\box0}
\def\pmbs#1{\setbox0=\hbox{$\scriptstyle #1$}%
  \kern-.0175em\copy0\kern-\wd0
  \kern.035em\copy0\kern-\wd0
  \kern-.0175em\raise.0303em\box0}

\def\be{\begin{equation}}
\def\ee{\end{equation}}
\def\bea{\begin{eqnarray}}
\def\eea{\end{eqnarray}}
\def\lb{\label}

\def\vec#1{\mbox{\boldmath$#1$}}

\def\gam{\gamma}
\def\d{\delta}
\def\eps{\epsilon}

\def\sig{\sigma}

\def\Sig{\Sigma}

\def\om{\omega}
\def\Om{\Omega}

\def\udot{\dot{u}}
\def\Udot{\dot{U}}

\def\cm{{\cal M}}

\def\bom{\mbox{\boldmath $\omega$}}

\def\bna{\mbox{\boldmath $\nabla$}}

\def\bu{\mbox{\boldmath $u$}}
\def\bv{\mbox{\boldmath $v$}}

\def\vece{\vec{e}}

\def\bfg{{\bf g}}
\def\hbfg{{\hat {\bf g}}}
\def\ptl{\partial}
\def\parb{\pmb{\partial}}

\def\la{\langle}
\def\ra{\rangle}

\def\hsp5{\hspace{5mm}}
\newcommand{\sfrac}[2]{{\textstyle{#1\over#2}}}
\def\case#1/#2{\textstyle\frac{#1}{#2}}
\def\ann{{Ann.~Phys.~(N.Y.)} }

\def\cmp{{Commun.~Math.~Phys.} }
\def\cqg{{Class.~Quantum~Grav.} }
\def\grg{{Gen.~Rel.~Grav.} }

\newcommand{\enl}{\\\hfill\rule{0pt}{0pt}}

\begin{document}

\title{\sc Conformal regularization of Einstein's field equations}
\author{\sc
 Niklas R\"ohr$^{1}$\thanks{Electronic address: {\tt Niklas.Rohr@kau.se}}\
,\ \ and Claes Uggla$^{1}$\thanks{Electronic address:
{\tt Claes.Uggla@kau.se}}\\
$^{1}${\small\em Department of Physics, University of Karlstad,}\\
{\small\em S-651 88 Karlstad, Sweden}}

\date{\normalsize{July 15, 2005}}
\maketitle

\begin{abstract}

To study asymptotic structures, we regularize Einstein's field
equations by means of conformal transformations. The conformal
factor is chosen so that it carries a dimensional scale that
captures crucial asymptotic features. By choosing a conformal
orthonormal frame we obtain a coupled system of differential
equations for a set of dimensionless variables, associated with
the conformal dimensionless metric, where the variables describe
ratios with respect to the chosen asymptotic scale structure. As
examples, we describe some explicit choices of conformal factors
and coordinates appropriate for the situation of a timelike
congruence approaching a singularity. One choice is shown to just
slightly modify the so-called Hubble-normalized approach, and one
leads to dimensionless first order symmetric hyperbolic equations.
We also discuss differences and similarities with other conformal
approaches in the literature, as regards, e.g., isotropic
singularities.

\end{abstract}
\centerline{\bigskip\noindent PACS number(s): 04.20.-q, 98.80.Jk,
04.20.Dw, 04.20.Ha~\hfill 
}\vfill
\newpage

\section{Introduction}

The importance of conformal properties and causal structure in
connection with Einstein's field equations (EFEs) is well known,
particularly as regards asymptotic structure, see e.g.
\cite{book:penrin86,book:hawell73,book:frafri02} and references
therein. It is also well known that scale-invariant, self-similar
\cite{carcol99}, solutions act as important building blocks for
our understanding of the asymptotic properties of
non-scale-invariant solutions, see e.g. \cite{book:waiell97} and
references therein. Indeed, the latter feature motivated the
introduction of the so-called Hubble-normalized dynamical systems
formulation of EFEs \cite{uggetal04} (henceforth denoted as UEWE),
which yielded regularized field equations in the neighborhood of
generic singularities and progress as regards our understanding,
and ability to numerically handle \cite{andetal05,gar04},
asymptotic dynamics towards such singularities.

The purpose of this paper is to provide a framework that naturally
captures all the above aspects. We thus combine the conformal and
dynamical systems approaches to EFEs into a common geometric
instrument; conformal transformations are used to obtain
scale-invariant, and thus dimensionless, regularized formulations
of EFEs that naturally incorporate key asymptotic causal
structures. The goal is to unravel features of the solution space
and properties of solutions of EFEs. The relationship between
conformal and asymptotic structures provides a systematic
geometric foundation for finding suitable geometrically
interpretable variables for dynamical systems analysis. This is to
be contrasted with the more or less random ad hoc choices of
variables that characterize the history of dynamical systems
studies in general relativity.

The outline of the paper is as follows: In Sec. 2 we discuss
dimensionality under scale-transformations and use this to
restrict the choice of conformal factor and frame; subsequently we
give a set of conformal, dimensionless, field equations. As an
example, we address asymptotic temporal structures associated with
singularity formation in Sec. 3. We thus consider a timelike
congruence and make a 1+3 split of our variables and equations. We
then give some examples of useful conformal and temporal gauge
choices: one that geometrizes and slightly modifies the
Hubble-normalized approach used in e.g. UEWE and \cite{gar04}, and
one that can be extended and modified to a dimensionless
autonomous first order symmetric hyperbolic system. In Sec. 4 we
conclude with a discussion and a comparison with other work; we
also outline our underlying philosophy.

\section{Scales and conformal transformations}

\subsection{Scales and dimensions}

Consider a spacetime $(M,\hbfg)$ where $M$ is a suitably smooth
4-dimensional manifold and $\hbfg$ is the physical Lorentzian
metric field with signature $(-,+,+,+)$. We use units $c=1=8\pi G$
so that all geometric properties can be dimensionally expressed in
terms of a length scale. Let $\ell$ be the unit of length, then
each physical geometric field $\Phi$ transforms under a scale
transformation $\ell'=S\ell$, where $S=const$, like
$\Phi'=S^q\,\Phi$, where $q$ defines the geometrical object's
dimension, see \cite{ear74} which we refer to for further
discussion.

General relativity is characterized by general coordinate
covariance; coordinates $x^\mu$ $(\mu=0,1,2,3)$ are to be regarded
as just labels for different spacetime events. Hence, in general,
they do not carry any physical significance and thus it is natural
to regard them as dimensionless, i.e., they have $q=0$. This is in
contrast to the spacetime interval $ds^2$ which describes an
invariant physical property and naturally have a weight $q=2$.
Since $ds^2 = g_{\mu\nu}\,dx^\mu\,dx^\nu$ it follows that
$g_{\mu\nu}$ has $q=2$. Hence one has to take into account the
specific positioning of indices for a geometric object when
considering its dimensional weight; for example, in the case of
the energy-momentum tensor: ${\hat T}^{\mu\nu}$, ${\hat
T}^\mu{}_\nu$, ${\hat T}_{\mu\nu}$ have dimensions $q=-4$, $q=-2$,
$q=0$, respectively.

Although coordinates are not dimensional in general, it is natural
to assign them dimensional weight when they express invariant
physical properties that occur in special cases, which tend to
dominate the literature. For example, the radial area coordinate
when one has spherical symmetry; proper time or length along a
given timelike or spacelike congruence, respectively; the affine
parameter along a geodesic congruence---these quantities all
naturally carry weight $q=1$. Another important special case is
weak gravity where one has a Minkowski background. This background
is preferably expressed in Minkowski coordinates with natural
weight $q=1$, since they constitute affine parameters of geodesics
and thus express invariant features in the Minkowski spacetime.

For computational purposes, one needs to specify a frame and use
components. This introduces an additional element since it is
quite natural to assign different dimensions to different choices
of frames. There are three types of frames that make dimensional
counting particularly easy for the components of geometric
objects: (i) Coordinate frames, since the coordinates in general
are to be regarded as dimensionless. (ii) Orthonormal frames
(ONF); in this case it is natural to regard the constant metric
coefficients as dimensionless and instead let the orthonormal
one-forms carry dimension 1, while the dual vector fields have
$q=-1$. This yields that connection components have $q=-1$ while
ONF curvature components have $q=-2$. The ONF approach was taken
as the starting point in UEWE, which we refer to for further
discussions about dimensions in ONF contexts. (iii) The third
choice is the one that is going to be explored in this
paper---conformal ONF where the conformal factor carries the
dimension.

We define a conformal ONF by
\be \lb{defconfon} \hbfg = \Psi^2\,\eta_{ab}\,\bom^a\,\bom^b =
\Psi^2\,\bfg\, , \ee
where $\eta_{ab}$ are constants ($a,b=0,1,2,3$), and where
$\bfg=\eta_{ab}\,\bom^a\,\bom^b$ is an `unphysical' metric
expressed in an ONF. The conformal factor $\Psi>0$, which is a
function of the spacetime coordinates, is chosen so that it has
dimension length, i.e. $q=1$, so that the metric $\bfg$ (and the
one-forms $\bom^a$ and their dual frame vectors $\vece_a$) becomes
dimensionless.

Since $\Psi$ carries the dimensional scale, it follows that
everything is compared with this scale in terms of dimensionless
ratios. However, there exists no overall preferred global choice
of $\Psi$ suitable for all possible situations; this is
reminiscent to the coordinate issue --- there exists no global
coordinate choice for a general spacetime either. Precisely as in
the coordinate case, one has to patch together a complete
spacetime from regions where one has used different conformal
factors. Instead of aiming for some global conformal choice,
$\Psi$ is to be adapted to the particular local feature one is
interested in. This can be some preferred structure associated
with asymptotic features, or a structure associated with special
initial or boundary conditions or symmetries. Useful candidates
are obtained from suitable functions of dimensional coordinates or
scalars, or quantities that preserve the defining structure of a
special class of spacetimes. Examples are e.g. $\Psi={\hat t}$,
where ${\hat t}$ is proper time along a timelike congruence; $\Psi
\propto \theta^{-1}$, where $\theta$ is the expansion of a null or
timelike congruence ($q=-1$ for $\theta$); $\Psi=r$, where $r$ is
the radial area coordinate in the case of spherical symmetry. We
will use $\Psi = {\hat t}, \Psi \propto \theta^{-1}$, associated
with a timelike congruence, as examples below in the context of
asymptotic temporal properties toward spacetime singularities.

\subsection{Conformal transformations and dimensionless field equations}

Let us begin with some notation and definitions. Consider some
arbitrary metric $\bfg$ in some basis of vector fields $\vece_a$,
with a dual basis of one-forms $\{\mbox{\boldmath $\omega$}_a\}$,
i.e. $\langle\,\mbox{\boldmath $\omega$}^{a},\,{\bf
e}_{b}\,\rangle = \delta^{a}{}_{b}$. Let us further introduce a
connection $\bna_b {\bf e}_{a} = \Gamma^{c}\!_{ab}\,{\bf e}_{c}\,
,\, \Gamma^{c}\!_{ab} = \langle\, \bom^c,\,\bna_b {\bf
e}_{a}\,\rangle$, where $\bna_b := \bna_{\vece_b}$. We then assume
that the connection is (i) torsion-free and (ii) metric:
\be ({\rm i}) \,\,\, \bna_{\bu}\,{\bv} - \bna_{\bv}\,{\bu} -
[{\bu},{\bv}] = 0\, ,\qquad ({\rm ii}) \,\,\, \bna {\bf g}=0\,
,\ee
where ${\bu}$ and ${\bv}$ are two arbitrary vectors and
$\bna_{\bu}=u^a\bna_a$. The components of the metric in the basis
$\{ {\bf e}_a\}$ are given by $g_{ab} = {\bf g}({\bf e}_a,{\bf
e}_b)$. Commutation functions $c^c{}_{ab}$ are defined by
\be
 [{\bf e}_a,{\bf e}_b] = c^c{}_{ab}{\bf e}_c\, .
\ee
The curvature operator, defined by ${\cal R}(\bu,\bv) =
[\bna_{\bu},\bna_{\bv}] - \bna_{[\bu,\bv]}$, yields the Riemann
curvature tensor expressed in components as $R^a{}_{bcd} =
\langle\,\bom^a,{\cal R}({\bf e}_c,{\bf e}_d){\bf e}_b\,\rangle$,
while the Ricci curvature tensor and scalar, and the Einstein
tensor, are defined by $R_{bd} = R_{db} = R^a\!_{bad}\, ,\, R =
R^{ab}\!_{ab}\, ,\, G^a{}_b = R^a{}_b -\sfrac{1}{2}\,g^a{}_b\,R$.
The components of the connection and Riemann and Ricci curvature
are
\bea \lb{eq:con}\Gamma_{abc} &=& -\sfrac{1}{2}\left[\, {\bf
e}_a(g_{bc}) - {\bf e}_b(g_{ca}) - {\bf e}_c(g_{ab}) + c_{abc} +
c_{bca} -
c_{cab}\,\right] \\
R^a{}_{bcd} &=& 2{\bf e}_{[c }\Gamma^a{}_{|b|d]} +
2\Gamma^a{}_{f[c}\,\Gamma^f{}_{|b|d]} -
\Gamma^a{}_{bf}\,c^f{}_{cd} \\
\lb{eq:ric} R_{ab} &=& 2{\bf e}_{[c }\Gamma^c{}_{|a|b]} +
\Gamma^c{}_{dc}\,\Gamma^d{}_{ab} -
\Gamma^c{}_{ad}\,\Gamma^d{}_{bc}\, ,\eea
where $\Gamma_{abc} = g_{ad}\Gamma^d{}_{bc}$. It follows that
$R_{abcd} = -\,R_{abdc}$, $R_{abcd} = -\,R_{bacd}$, $R_{abcd} =
R_{cdab}$, $R^a{}_{[bcd]} = 0$, $\bna_{[e}\,R^a{}_{|b|cd]}=0$,
where the two last relations are the cyclic and Bianchi
identities, respectively. See e.g. \cite{book:grav73} for further
discussion.

Let us now consider a `physical'  metric $\hbfg$ conformally
related to an `unphysical' metric $\bfg$ according to
\be \hbfg = {\hat g}_{ab}\,\bom^a\,\bom^b =
\Psi^2\,g_{ab}\,\bom^a\,\bom^b = \Psi^2\,\bfg\, , \ee
where $\bom^a$ is a dual basis of one-forms to the basis
$\vece_a$: $\langle\,\mbox{\boldmath $\omega$}^{a},\,{\bf
e}_{b}\,\rangle = \delta^{a}{}_{b}$. Then the connection and Ricci
tensor of $\hbfg$ are related to the connection and Ricci tensor
of $\bfg$ according to (easily derivable from Eqs. (\ref{eq:con})
and (\ref{eq:ric}); or see e.g. \cite{book:exactsol03})
\bea \hat{\Gamma }^a{}_{bc} &=& \Gamma^a{}_{bc} +
2\delta^a{}_{(b}\,r_{c)} -
g_{bc}\,r^a \\
\lb{confric} \hat{R}_{ab} &=& R_{ab} -2\bna_{(a} r_{b)} + 2r_a r_b
- g_{ab}(\bna_c r^c + 2r^2)\, , \eea
in the basis $\vece_a$, where
\be r_a := \frac{\vece_a\,\Psi}{\Psi}\, , \ee
and $r^a=g^{ab}\,r_b$, $r^2=g^{ab}\,r_a\,r_b$ and
$\bna_a\,r_b={\bf e}_a\,r_b - \Gamma^c{}_{ba}\,r_c$.
We now express the unphysical metric $g_{ab}$ in an ONF so that
$g_{ab}=\eta_{ab}=\text{diag}[-1,1,1,1]$, and thus the physical
metric is given in a conformal ONF: ${\hat g}_{ab} =
\Psi^2\,\eta_{ab}$, see Eq.~(\ref{defconfon}).

We take the frame variables $e_a{}^\mu$, defined by
\be\vece_a = e_a{}^\mu\ptl/\ptl x^\mu\, , \ee
and the commutator functions $c^a{}_{bc}$ as our basic variables,
possibly supplemented by $r_a$; we will give some examples in Sec.
3.

The governing dimensionless equations are the commutator
equations, the Jacobi identities for $\vece_a$, and EFEs, which in
the conformal ONF, \{$\vece_a$\}, are given by
\bea \lb{comeq} 2{\bf e}_{[a}\,e_{b]}{}^{\mu} &=&
c^c{}_{ab}\,e_c{}^{\mu} \\
\lb{jaceq} {\bf e}_{[a}\,c^d{}_{bc]} &=& c^d{}_{e[a}\,c^e{}_{bc]} \\
\lb{fieldeq} \hat{R}_{ab} &=&
T_{ab}-\sfrac{1}{2}\eta_{ab}\,\eta^{cd}\,T_{cd} \ , \eea
where $T_{ab}$ are the dimensionless components of the
energy-momentum tensor in the conformal ONF (recall that the
energy-momentum tensor ${\hat T}_{\mu\nu}$ is dimensionless, i.e.
$q=0$, and hence it is possible to make the identification ${\hat
T}_{ab} = T_{ab}$ in the conformal ONF), and where now
\bea \hat{R}_{ab} &=& R_{ab} + U_{ab}\, ,\quad U_{ab} :=
-2\bna_{(a}\,r_{b)} + 2r_a \,r_b  - \eta_{ab}(\bna_c\,r^c +2r^2)
\\
\lb{gam} \Gamma^a{}_{bc} &=& -\sfrac{1}{2}\eta^{ad}\left[\,
\eta_{ed}\,c^e{}_{bc} + 2\eta_{e(b}\,c^e\,_{c)d}\,\right]\, . \eea
Note that $2\Gamma_{a[bc]} = -c_{abc}$ and $\Gamma_{a(bc)} =
-c_{(bc)a}$; one can thus use $\Gamma^a{}_{bc}$ as variables
instead of $c^a{}_{bc}$.

When one has a non-trivial matter source the above equations have
to be supplemented with appropriate matter equations, however, one
always have local energy-momentum conservation: ${\hat
\bna_a}{\hat T}^{ab}=0$. Recall that ${\hat T}^{\mu\nu}$ has
dimension $q=-4$. In the conformal ONF we have: ${\hat T}^{ab} =
{\hat g}^{ac}\,{\hat g}^{bd}\,{\hat T}_{cd} =
\Psi^{-4}\,\eta^{ac}\,\eta^{bd}\,T_{cd}=\Psi^{-4}\,T^{ab}$, where
we have defined the dimensionless object
$T^{ab}:=\eta^{ac}\,\eta^{bd}\,T_{cd}$. This yields (see
\cite{wu81});
\be {\bf e}_b\,T^{ab} + \Gamma^a{}_{db}\,T^{db} +
\Gamma^b{}_{db}\,T^{ad} + 2r_b\,T^{ab} - r^a\,\eta_{cd}\,T^{cd} =
0\, . \ee

It is of key importance to note that the equations for the
dimensional variable $\Psi$, $\vece_a\,\Psi =\Psi\,r_a$, decouple
from the above dimensionless equations (this is seen explicitly,
but also follows directly from dimensional reasons). The
components of $r_a$ are either given functions of coordinates or
functions of the dimensionless state space variables that depend
on what type of choice of $\Psi$ one has made, examples will be
given below. This means that $\Psi$ can be obtained after one has
solved the essential dimensionless equations, and thus $\Psi$
itself plays a `passive' subsidiary role. It is $\Psi$ that
carries the scale that typically asymptotically leads to that
dimensional quantities blow up. Since this scale now has been
factored out of the problem, leaving an asymptotically regularized
dimensionless system, this drastically simplifies an asymptotic
analysis. Once an asymptotic analysis has been accomplished for
the essential dimensionless equations, the result can be used for
a relatively simple asymptotic analysis of the decoupled equations
for $\Psi$, thus yielding a complete physical result. This
geometric splitting of the problem into more easily handled
problems is the conformal ONF approach's main advantage.

It is of interest to relate the present variables to those that
one uses in the ONF approach. In the latter approach one uses a
basis so that $\hbfg = \eta_{ab}\,{\hat \bom}^a\,{\hat \bom}^b$
and an associated dual basis ${\hat \vece}_a$, $\langle\,{\hat
\bom}^a,\,{\hat \vece}_b\,\rangle = \delta^{a}{}_{b}$. The
variables in this approach are the frame variables ${\hat
e}_a{}^\mu$, defined by ${\hat \vece}_a = {\hat
e}_a{}^\mu\ptl/\ptl x^\mu$, and the commutator variables ${\hat
c}^a{}_{bc}$, defined by $[{\hat \vece}_b,{\hat \vece}_c] = {\hat
c}^a{}_{bc}{\hat \vece}_a$. These variables are related to the
present ones as follows:
\bea \lb{ecomprel}{\hat e}_a{}^\mu &=& \Psi^{-1}\,e_a{}^\mu\, ;
\qquad\qquad\qquad\quad
e_a{}^\mu =  \Psi\,{\hat e}_a{}^\mu\\
\lb{strucfrel}{\hat c}^a{}_{bc} &=& \Psi^{-1}\,\left(\,c^a{}_{bc}
+ \delta^a{}_{[b}\,r_{c]}\,\right)\, ;\quad c^a{}_{bc} =
\Psi\,{\hat c}^a{}_{bc} - \delta^a{}_{[b}\,r_{c]}\, ,\eea
where ${\hat c}^a{}_{bc} = \langle\, {\hat \bom}^a,[{\hat
\vece}_b,{\hat \vece}_c]\,\rangle$ and $c^a{}_{bc} = \langle\,
\bom^a,[\vece_b,\vece_c]\,\rangle$, i.e., the above equations are
not tensor equations; instead they relate the components of the
frame variables and the commutator functions of two conformally
related sets of basis vector fields. The above relationships
explicitly show that $e_a{}^\mu$ and $c^a{}_{bc}$ are
dimensionless if $\Psi$ has dimensional weight $q=1$, and that
everything is measured with respect to the scale carried by
$\Psi$.

To make this more concrete we will consider a timelike congruence
and give some examples of conformal and temporal gauge choices.

\section{The 1 + 3 conformally orthonormal approach}

\subsection{1+3 decomposition}

We here adapt our formalism to a timelike reference congruence. We
therefore choose a time coordinate along the congruence and align
one of the basis vectors tangentially to it; this allows us to
make a 1+3 split of the variables.

The main application in this paper is the use of conformal
regularization towards a generic singularity. This means that
there is a close connection with UEWE. Unfortunately the notation
in UEWE is not adapted to the conformal formalism at all, which
suggests that it perhaps would be best to use new notation that is
naturally associated with the conformal approach. Nevertheless, in
this paper we adopt a notation that follows that of UEWE as
closely as possible, since this emphasizes the close connection
and simplifies a comparison between UEWE and the present work,
even though this leads to some awkwardness.

In UEWE the starting point was the ONF formalism associated with
the physical metric. Since conformal transformations were not
discussed in UEWE the frame vectors did not have hats. We
therefore now drop the hats on the ONF vectors, i.e., $\{{\hat
\vece}_a\} \rightarrow \{\vece_a\}$. This causes a problem for the
conformal ONF vectors, however, these vectors correspond to the
Hubble-normalized vectors $\parb_a$ in UEWE; we thus rename the
conformal ONF vectors according to $\{\vece_a\} \rightarrow
\parb_a$, but in contrast to UEWE (associated with $\Psi=H^{-1}$ where
$H$ is the Hubble variable), $\Psi$ is now any function with
$q=1$. A 1+3 split (in contrast to the 3+1 split done in UEWE; see
e.g. \cite{hveugg97}, and references therein) of the ONF and
conformal ONF vectors and variables yields:
\bea \lb{13ONfr} \vece_{0} &=& M^{-1}\,\ptl_{t}\,, \qquad
\vece_{\alpha} = e_{\alpha}{}^{i}(M_{i}\,\ptl_{t}+\ptl_{i}) \\
\lb{13cONfr} \parb_{0} &=& \cm^{-1}\,\ptl_{t} \ , \qquad
\parb_{\alpha} = E_{\alpha}{}^{i}(M_{i}\,\ptl_{t}+\ptl_{i}) \ ,
\qquad \alpha = 1,2,3 \ ; \quad i = 1,2,3\, , \eea
where $\vece_{0}$ ($\parb_0$) is the future-directed tangent to
the physical (unphysical conformal) timelike reference congruence;
$M$ ($\cm$) is the physical (unphysical conformal) threading lapse
function, and $M_{i}$ the dimensionless (assuming dimensionless
coordinates) threading shift one-form.

The lapse, $M=\Psi\,\cm ;\, \cm = \Psi^{-1}\,M$, and the
dimensionless shift vector are associated with gauge freedom while
$e_\alpha{}^i$ and $E_\alpha{}^i$ are regarded as dynamical
variables, related by
\be \lb{ecomprel} e_\alpha{}^i = \Psi^{-1}\,E_\alpha{}^i\, ;
\qquad E_\alpha{}^i =  \Psi\,e_\alpha{}^i\, .\ee

The commutators are decomposed according to:
\bea \lb{dcomts0} [\,\vece_{0}, \vece_{\alpha}\,] & = &
\dot{u}_{\alpha}\,\vece_{0} - [\,H\,\d_{\alpha}{}^{\beta} +
\sig_{\alpha}{}^{\beta} +
\eps_{\alpha}{}^{\beta}{}_{\gam}\,(\omega^{\gam}+\Omega^{\gam})\,]\,
\vece_{\beta} \\
\lb{dcomtsa} [\,\vece_{\alpha}, \vece_{\beta}\,] & = &
2\eps_{\alpha\beta\gam}\,\omega^{\gam}\,\vece_{0} +
(2a_{[\alpha}\,\d_{\beta]}{}^{\gam} +
\eps_{\alpha\beta\delta}\,n^{\delta\gam})\,\vece_{\gam} \\
\lb{dcomts0} [\,\parb_{0}, \parb_{\alpha}\,] & = &
\dot{U}_{\alpha}\,\parb_{0} - [\,{\cal H}\,\d_{\alpha}{}^{\beta} +
\Sig_{\alpha}{}^{\beta} +
\eps_{\alpha}{}^{\beta}{}_{\gam}\,(W^{\gam}+R^{\gam})\,]\,
\parb_{\beta} \\
\lb{ccomtsa} [\,\parb_{\alpha}, \parb_{\beta}\,] & = &
2\eps_{\alpha\beta\gam}\,W^{\gam}\,\parb_{0} +
(2A_{[\alpha}\,\d_{\beta]}{}^{\gam} +
\eps_{\alpha\beta\delta}\,N^{\delta\gam})\,\parb_{\gam}\, ,\eea
where the above decomposition imply the following definitions:
\bea \lb{decomp1} H & = & -\sfrac{1}{3} {\hat
c}^\alpha{}_{0\alpha}\, ,\quad \sigma_{\alpha\beta}  = -{\hat
c}^\gam{}_{0\la\alpha}\d_{\beta\ra\gam}\, ,\quad \udot_\alpha =
{\hat c}^0{}_{0 \alpha }\, ,
\\
\lb{decomp2}\omega_\alpha & = &
\sfrac{1}{4}\epsilon_\alpha{}^{\beta\gam}\,{\hat
c}^0{}_{\beta\gam}\, ,\quad \omega_\alpha + \Omega_\alpha =
\sfrac{1}{2}\epsilon_{\alpha\beta}{}^\gam\,{\hat
c}^\beta{}_{0\gam}\, ,\quad n^{\alpha\beta} =
\sfrac{1}{2}\epsilon^{\mu\nu(\alpha}\,{\hat
c}^{\beta)}{}_{\mu\nu}\, ,\quad
a_\alpha  =  \sfrac{1}{2}{\hat c}^\beta{}_{\alpha\beta} \\
\lb{cecomp1}{\cal H} & = & -\sfrac{1}{3} c^\alpha{}_{0\alpha}\,
,\quad \Sigma_{\alpha\beta}  =
-c^\gam{}_{0\la\alpha}\d_{\beta\ra\gam}\, ,\quad \Udot_\alpha =
c^0{}_{0 \alpha }\, ,
\\
\lb{cecomp2}W_\alpha & = &
\sfrac{1}{4}\epsilon_\alpha{}^{\beta\gam}\,c^0{}_{\beta\gam}\,
,\quad W_\alpha + R_\alpha =
\sfrac{1}{2}\epsilon_{\alpha\beta}{}^\gam\,c^\beta{}_{0\gam}\,
,\quad N^{\alpha\beta} =
\sfrac{1}{2}\epsilon^{\mu\nu(\alpha}\,c^{\beta)}{}_{\mu\nu}\,
,\quad A_\alpha  =  \sfrac{1}{2}c^\beta{}_{\alpha\beta}\ , \eea
where $\la\,\,\ra$ represents trace free symmetrization. Here
$H=\frac{1}{3}\theta$ is the Hubble variable, and $\theta$ the
expansion; $\sig_{\alpha\beta}$ the shear; $\udot_\alpha$ the
acceleration; $\omega_\alpha$ the rotation; $\Omega_\alpha$ the
Fermi rotation---all quantities are associated with the congruence
of which $\vece_0$ is the tangent vector field; $n^{\alpha\beta}$,
$a_\alpha$ are spatial commutator functions, which describe the
three-curvature when $\omega_\alpha=0$; for a more detailed
description, see e.g. \cite{book:waiell97}, \cite{hveugg97}.
Analogous interpretations hold for the conformal quantities. In
the above formulas we have adhered to the conventions used in
\cite{book:ellels99}.

Eqs.~(\ref{strucfrel}), (\ref{decomp1})-(\ref{cecomp2}) yield the
following relationship between the ONF and conformal ONF
commutator function variables:
\begin{alignat}{3}\lb{dim1} {\cal H} & =
\Psi H -r_{0}\, ,& \qquad  \Sig_{\alpha\beta} &=
\Psi\sig_{\alpha\beta}\, ,\\
\lb{dim2}\Udot_\alpha & = \Psi\udot_\alpha\ -
r_{\alpha}\, ,& \qquad W_{\alpha} & = \Psi\om_{\alpha}\, ,\\
\lb{dim3} N^{\alpha\beta} & = \Psi n^{\alpha\beta}\, ,& \qquad
A_{\alpha} & = \Psi a_{\alpha} +
r_{\alpha}\, , \\
\lb{dim3} W^\alpha + R^\alpha  & = \Psi(\om^\alpha+\Om^\alpha)\,.&
\qquad  &
\end{alignat}

Let us now focus on the conformal ONF approach. From the above
definitions and Eq.~(\ref{gam}) it follows that the 1 + 3 splitted
connection components of the conformal metric are given by
\bea\lb{decomp_conn}\Gamma_{\alpha00} & = & \Udot_{\alpha}\,
,\qquad\qquad
 \Gamma_{\alpha0\beta} = {\cal H}\d_{\alpha\beta} +
 \Sig_{\alpha\beta} - \epsilon_{\alpha\beta\gamma}W^{\gamma}\,,\\
 \Gamma_{\alpha\beta0} & = &
 \epsilon_{\alpha\beta\gamma}R^{\gamma}\, ,\qquad
 \Gamma_{\alpha\beta\gamma} = 2A_{[\alpha}\d_{\beta]\gamma} +
 \epsilon_{\gamma\d[\alpha}N^{\d}{}_{\beta]} +
 \sfrac{1}{2}\epsilon_{\alpha\beta\d}N^{\d}{}_{\gamma}\, .\eea
Instead of referring to ${\cal H},
\Sig_{\alpha\beta},\Udot_\alpha, W_\alpha, R_\alpha, A_\alpha,
N^{\alpha\beta}$ as commutator function variables, one may refer
to them as connection variables, since they describe
$\Gamma^a{}_{bc}$ as well as $c^a{}_{bc}$.

The commutator equations can be written succinctly as follows:
\bea \lb{id0a} 0 & = & (\parb_{\alpha}+ \Udot_{\alpha})\parb_{0} -
(\d_{\alpha}{}^{\beta}\,\parb_{0}
- F_{\alpha}{}^{\beta})\,\parb_{\beta} \\
\lb{idab} 0 & = & 2W_{\alpha}\,\parb_{0} -
\vec{C}_{\alpha}{}^{\beta}\,\parb_{\beta}\, , \eea
where
\bea F_{\alpha}{}^{\beta} & := & c^{\beta}{}_{0\alpha} =
-{\cal{H}}\,\d_{\alpha}{}^{\beta} - \Sig_{\alpha}{}^{\beta}
- \eps_{\alpha}{}^{\beta}{}_{\gam}\,(W^{\gam}+R^{\gam}) \\
\vec{C}_{\alpha}{}^{\beta} & := & \eps_{\alpha}{}^{\gam\beta}\,
(\parb_{\gam}-A_{\gam}) - N_{\alpha}{}^{\beta} \ , \eea
and this suggests that the equations can be written concisely
using the above notation.

It is natural to divide the equations into gauge equations and
dynamical equations, and to further subdivide the latter into
evolution equations and constraints (if the temporal frame
derivative, $\parb_0$, does not appear in a dynamical equation we
refer to it as a constraint equation, even though the spatial
frame derivatives $\parb_\alpha$ contain the partial time
derivative $\ptl_t$):\enl

\noindent {\em Gauge equations\/}:
\bea \lb{dlM} \parb_{0}{\cal M}_{\alpha} & = &
F_{\alpha}{}^{\beta}\,{\cal M}_{\beta}
+ (\parb_{\alpha} + \Udot_{\alpha}){\cal M}^{-1}\, \\
\lb{dl13comssw} 0 & = & \vec{C}_{\alpha}{}^{\beta}\,{\cal
M}_{\beta} - 2{\cal M}^{-1}\,W_{\alpha} \ . \eea

\noindent {\em Evolution equations\/}:
\bea \lb{dl13comts} \parb_{0}E_{\alpha}{}^{i} & = &
F_{\alpha}{}^{\beta}\,E_{\beta}{}^{i}\, \\
\lb{dhdot} \parb_{0}{\cal H} & = & -{\cal H}^2
-\sfrac{1}{3}\,\Sig_{\alpha\beta}\Sig^{\alpha\beta} +
\sfrac{2}{3}\,W^2 + \sfrac{1}{3}\,(\parb_{\alpha}+ \Udot_{\alpha}
- 2A_\alpha)\,\Udot^\alpha - \sfrac{1}{6}(T_{00} +
T^\alpha{}_\alpha) +\sfrac{1}{3}U_{00}\,\\
\lb{dlsigdot} \parb_{0}\Sig_{\alpha\beta} & = & -3{\cal
H}\,\Sig_{\alpha\beta} -
\eps^{\gam\delta}{}_{\la\alpha}\,(2\Sig_{\beta\ra\gam}\,R_{\d}\, -
N_{\beta\ra\gam}\,\Udot_{\delta}) - 2W_{\la\alpha}\,R_{\beta\ra} +
(\parb_{\la\alpha} + A_{\la\alpha} + \Udot_{\la\alpha})\,
\Udot_{\beta\ra}
\nonumber  \\
& & \hsp5 \ - \,{}^{3}{\cal S}_{\alpha\beta}
 + T_{\la\alpha\beta\ra}- U_{\la\alpha\beta\ra} \, \\
\lb{dlwdot} \parb_{0}W_{\alpha} & = &
-(3{\cal{H}}\d_{\alpha}{}^{\beta} +
F_{\alpha}{}^{\beta})\,W_{\beta}
+ \sfrac{1}{2}\,\vec{C}_{\alpha}{}^{\beta}\,\Udot_{\beta}\, \\
\lb{dladot} \parb_{0}A_{\alpha} & = &
F_{\alpha}{}^{\beta}\,A_{\beta} - \sfrac{1}{2}\,(\parb_{\beta}+
\Udot_{\beta})(\,
3{\cal{H}}\,\d_{\alpha}{}^{\beta}+F_{\alpha}{}^{\beta}\,)\, \\
\lb{dlndot} \parb_{0}N^{\alpha\beta} & = &
-(3{\cal{H}}\,\d_{\gam}{}^{(\alpha}+2F_{\gam}{}^{(\alpha}{})\,N^{{\beta}){\gam}}
+ (\parb_{\gamma}+ \Udot_{\gamma})\eps^{\gamma\delta
(\alpha}F_{\delta}{}^{\beta)} \ . \eea
\noindent {\em Constraint equations\/}:
\bea \lb{dl13com}
0 & = & \vec{C}_{\alpha}{}^{\beta}\,E_{\beta}{}^{i}\, \\
\lb{dlgauss} 0 & = & 3{\cal H}^2 -
\sfrac{1}{2}\,\Sig_{\alpha\beta}\Sig^{\alpha\beta} + W^{2} -
2W_{\alpha}R^{\alpha} + \sfrac{1}{2}\,{}^{3}{\cal R}
- T_{00} + \sfrac{1}{2}(U_{00} + U^{\alpha}{}_{\alpha})\, \\
\lb{dlcodacci} 0 & = & -2\parb_\alpha {\cal H} +
(\parb_{\beta}-3A_{\beta})\Sig_{\alpha}{}^{\beta} +
\eps_{\alpha}{}^{\beta\gam}\,(\Sig_{\beta}{}^{\delta}\,
N_{\delta\gam} + 2\Udot_{\beta}\,W_{\gam}) +
\vec{C}_{\alpha}{}^{\beta}\,W_{\beta} - T_{0\alpha} +
U_{0\alpha}\, \\
\lb{dljacobi1} 0 & = &
(\parb_{\beta}-2A_{\beta})\,N_{\alpha}{}^{\beta} +
\eps_\alpha{}^{\beta\gamma}\parb_{\beta}A_{\gamma} -
2F_{\alpha}{}^{\beta}\,W_{\beta}\, \\
\lb{dljacobi2} 0 & = & (\parb_\alpha-\Udot_\alpha
-2A_{\alpha})\,W^{\alpha}\, , \eea
where
\bea {}^{3}{\cal S}_{\alpha\beta} & = & \parb_{\la\alpha}\,
A_{\beta\ra} - (\parb_\gam -
2A_\gam)N_{\d\la\alpha}\eps_{\beta\ra}{}^{\gam\d} +
B_{\la\alpha\beta\ra}\, \\
{}^{3}{\cal R} & = & 4\parb_{\alpha}A^{\alpha} - 6A^2 -
\sfrac{1}{2}B_\alpha{}^\alpha \, \\
B_{\alpha\beta} & = & 2N_{\alpha}{}^\gam\,N_{\gam\beta} -
N_\gam{}^\gam\,N_{\alpha\beta}\, \\
\lb{U00} U_{00} & = & - 3(\parb_0 + {\cal H})r_0 +
\left[\d^{\beta\gam}(\parb_{\beta} + 2r_{\beta}) + 3\Udot^{\gam}
-2A^{\gam} \right]r_{\gam}\,
 \\
\lb{U0alpha} U_{0\alpha} & = & -2(\parb_{\alpha}-r_{\alpha})r_0 +
2({\cal H}\d_{\alpha}{}^{\beta}+\Sigma_{\alpha}{}^{\beta}+
\epsilon_{\alpha}{}^{\beta\gamma}W_{\gamma})r_{\beta}\, \\
\lb{Utraceless} U_{\la\alpha\beta\ra} & = &
2\left[\Sig_{\alpha\beta}r_0 - (\parb_{\la\alpha} +
A_{\la\alpha})r_{\beta\ra}
-\eps^{\gam}{}_{\d\la\alpha}N^{\d}{}_{\beta\ra}r_{\gam} +
r_{\la\alpha}r_{\beta\ra} \right]\, \\
\lb{U00trace} U_{00} + U^{\alpha}{}_{\alpha} & = & 6(2{\cal H} +
r_0)r_0 - 2\left[\d^{\beta\gamma}(2\parb_{\beta} + r_{\beta}) -
4A^{\gamma} \right]r_{\gamma}\, , \eea
where we have used the notation $v_{\alpha}v^{\alpha} = v^{2}$. If
$M_\alpha=0=W_\alpha$, then ${}^{3}{\cal R}$ and ${}^{3}{\cal
S}_{\alpha\beta}$ are the curvature scalar and trace free part of
the Ricci tensor, respectively, of the conformal 3-metric.


A conformal 1+3 split of the equations for $T^{ab}$ yields (to
avoid clashes with the notation in UEWE we will refrain from
explicitly splitting $T^{ab}$ in terms of its irreducible parts,
but see the discussion below):
\bea (\parb_0 + 3{\cal H})\,T_{00} + {\cal H}\,T^{\beta}{}_{\beta}
- (\parb_{\beta} + 2\Udot_{\beta} - 2A_{\beta})\,T_0{}^{\beta}  +
\Sig_{\beta\gamma}\,T^{\beta\gamma} + C_0 & = & 0\, \\
(\parb_0 + 4{\cal H})\,T_{0\alpha} - \Udot_{\alpha}\,T_{00} -
A_{\alpha}\,T^{\beta}{}_{\beta} +
\Sig_{\alpha}{}^{\beta}\,T_{0\beta} - (\parb_{\beta} +
\Udot_{\beta}
- 3A_{\beta})\,T_{\alpha}{}^{\beta} && \\
+
\,\,\eps_{\alpha}{}^{\beta\gamma}\left[N_{\beta}{}^{\d}\,T_{\d\gamma}
- (W_{\gamma} - R_{\gamma})\,T_{0\beta}\right] + C_{\alpha} & = &
0\, ,\eea
where
\bea C_0 & := & (T_{00} + T^{\beta}{}_{\beta})\,r_0 -
2T_0{}^{\beta}\,r_{\beta}\, \\
C_{\alpha} & := & 2T_{0\alpha}\,r_0 - (T_{00} -
T^{\beta}{}_{\beta})\,r_{\alpha} -
2T_{\alpha}{}^{\beta}\,r_{\beta}\, . \eea

As done in e.g. UEWE, the energy-momentum tensor can be 1+3
splitted according to
\be {\hat T}^{ab} = {\hat \rho}\,{\hat u}^a\,{\hat u}^b + 2\,{\hat
u}^a\,{\hat q}^{b)} + {\hat p}\,{\hat h}^{ab} + {\hat \pi}^{ab}\,,
\ee
where ${\hat u}_a{\hat h}^{ab}=0$, ${\hat u}_a{\hat q}^a=0$,
${\hat \pi}^a{}_a=0$. In \cite{book:waiell97} and UEWE the
following normalization was introduced: $(\Omega, P, Q^\alpha,
\Pi^{\alpha\beta}) = ({\hat \rho},{\hat p}, {\hat q}^\alpha,{\hat
\pi}^{\alpha\beta})/(3H^2)$; the reason for this is that this
yields the standard definition of the important cosmological
dimensionless density parameter $\Omega$. However, from a
conformal geometric perspective it follows that if one wants to
use $H^{-1}$ as conformal factor, then the natural normalization
factor is $H^{-2}$. This then suggests the following new
definitions in the Hubble-normalization case: $(D, P, Q^\alpha,
\Pi^{\alpha\beta}) = ({\hat \rho},{\hat p}, {\hat q}^\alpha,{\hat
\pi}^{\alpha\beta})/H^2$, associated with the irreducible 1+3
decomposition of $T^{ab}$, and hence $\Omega = D/3$; alternatively
one may use $(\sqrt{3}H)^{-1}$ as conformal factor, but that
changes the conventions with respect to \cite{book:waiell97} and
UEWE as regards the connection variables.

\subsection{Examples of conformal and gauge choices}

Associated with a choice of a conformal normalization factor
$\Psi$, there exists a natural temporal gauge choice---the
conformal `proper time gauge', $\cm=1$, which in the 3+1 case $M_i
= 0 = W_\alpha$ reduces to the conformal Gauss gauge (see
\cite{book:frafri02} for a discussion about the use of conformal
Gauss coordinates to cover large spacetime domains), however, it
is of course not necessary to choose this gauge. Instead of a
general discussion about conformal, frame, and coordinate freedom,
we will focus on a few examples. Since the emphasis in this paper
is on the conformal approach, we will divide our discussion in
terms of conformal choices; we consider two such
choices---conformal Hubble-normalization and conformal proper time
normalization.

\subsubsection{Conformal Hubble-normalization}

The first example is given by
\be \Psi = H^{-1}\, , \ee
where $H=\frac{1}{3}\theta$ is the physical Hubble variable and
$\theta$ is the physical expansion, defined by $\theta = {\hat
\bna}_a\,{\hat u}^a$, where ${\hat {\bu}} = {\hat \vece}_0$ (in
the notation of Sec. 2). In this case we define the physical
deceleration parameter $q$ (not to be confused with the scale
weight $q$) and logarithmic spatial frame derivative $r_\alpha $
by
\be \lb{Hubblenorm}\parb_0 H = -(1+q)H\, ,\qquad \parb_\alpha H =
-r_\alpha\,H\, ,\ee
i.e., in terms of $r_a$ we have: $r_0=1+q = -\parb_0 H/H$ and
$r_\alpha = -\parb_\alpha H/H$, which combined with Eqs.
(\ref{U00})-(\ref{U00trace}) determine $U_{ab}$.

Eqs.~(\ref{dim1})-(\ref{dim3}) and the above definitions yield:
\be\lb{Hrel}{\cal H} = -q\, ,\qquad \Udot_\alpha = \Udot_\alpha^H
- r_{\alpha}\, ,\qquad A_{\alpha}= A_{\alpha}^H + r_{\alpha}\, ,
\ee
where $\Udot_\alpha^H:=\udot_\alpha/H$,
$A_\alpha^H:=a_{\alpha}/H$, while the other variables are just the
usual Hubble-normalized variables used in e.g. UEWE.

One can choose to let $q$ and $r_\alpha$ be determined by the
Raychadhuri equation (\ref{dhdot}) (the time derivative of $q$
drops out when ${\cal H} = -q, r_0 = 1+q$ are inserted in
(\ref{dhdot})) and the Codacci constraint (\ref{dlcodacci}) (the
spatial derivatives of $q$ drop out), respectively. However, since
$-q$ is just one of the connection variables in the conformal
formulation, it is quite natural to extend the normalized state
space to include $q$ and $r_\alpha$ (which is also connected to
the present formalism through its link to the gauge quantity
$\Udot_\alpha$) as independent variables, something which has been
found to be quite useful, see e.g. \cite{gar04} and
\cite{gargun05}.

Setting $\cm=1$ yields the separable volume gauge, see UEWE, and
if one in addition sets $M_\alpha=0=W_\alpha$ one obtains the
inverse mean curvature gauge, which in the present context can be
interpreted as the conformal Gauss gauge associated with
$\Psi=H^{-1}$, something which is also reflected in that the
congruence is conformally geodesic: $\Udot_\alpha=0$. This further
emphasizes the geometric nature of the present approach, and the
preference of using $\Udot_\alpha$ instead of $\Udot_\alpha^H$.

Note that with the current choice of conformal factor, and a
conformal Gauss coordinate choice, the present formulation reduces
to the `standard' Hubble-normalized formulation in the spatially
homogeneous (SH) case, since $\Udot_\alpha=0=r_\alpha$, i.e., {\em
the present formulation yields a natural geometric generalization
of the Hubble-normalized SH case\/}, discussed extensively in e.g.
\cite{book:waiell97}. For the same reason the present approach
reduces to that used in e.g. UEWE for the SH part of the so-called
silent boundary, where the attractor for generic singularities
resides (this is also the case for the subset associated with
isotropic singularities \cite{limetal04}). {\em Hence the
description of the attractor for a generic singularity in the
present geometric formulation is identical to that in UEWE---the
asymptotic regularization properties are generically identical\/}.

\subsubsection{Conformal proper time normalization}

The second example uses the physical proper time ${\hat t}$ along
a timelike reference congruence as the conformal factor:
\be \Psi={\hat t}\, .\ee
The time variable is subsequently reparametrized so that one
obtains a dimensionless time variable, $t$, according to
\be t= \ln({\hat t}/{\hat t}_0)\, ,\qquad  {\hat t}={\hat
t}_0\,e^t\, ,\ee
where ${\hat t}_0$ is some reference time; it follows that the new
time variable is just the conformal proper time, since $\cm=1$,
and hence $\parb_0=\ptl/\ptl_t$ and $r_a = (1,0,0,0)$. This leads
to
%
%
\be U_{00}=-3{\cal H}\, , \quad  U_{0\alpha}=2\Udot_{\alpha}\,
,\quad U_{\la\alpha\beta\ra}=2\Sig_{\alpha\beta}\, ,\quad
U_{00}+U^{\alpha}{}_{\alpha}= 6(2{\cal H} + 1)\, , \ee
and thus one obtains a first order {\em autonomous\/} system of
equations. This can be seen explicitly, but again this also
follows from dimensional reasons: ${\hat t}$ is the only varying
quantity that carries dimension and ${\hat t}$ can hence not
appear in the dimensionless equations, and therefore the same
holds for $t$; neither does the normalization affect the essential
first order structure of the usual dimensional ONF approach.

In \cite{fri98}, \cite{ellels99}, and \cite{elletal00} it was
shown that by extending the ONF approach to also include the
curvature tensor and the Bianchi identities one can obtain a first
order symmetric hyperbolic system, if one uses proper time along a
timelike congruence (this was shown for a perfect fluid with a
barotropic equation of state by using proper time along the fluid
congruence). It is of course also possible to extend the current
conformal ONF approach similarly, as will be discussed in the next
section. Since $\Psi={\hat t} = {\hat t}_0e^t$ does not modify the
principle parts of the equations of a curvature extended
formulation, we conclude that it is possible to extend the present
`minimal' formalism and obtain a dimensionless autonomous first
order symmetric hyperbolic system for the ${\hat t}$-normalized
equations; incidentally, this system is of course well-posed.

Let us now for simplicity specialize the temporal reference
congruence to be non-rotating, $M_\alpha=0=W_\alpha$, so that we
obtain proper time normalized equations and a conformal Gauss
coordinate system, for which $\Udot_\alpha=0$ and hence
$U_{0\alpha}=0$. Moreover, let us consider a generic initial
spacelike singularity and let us specialize the time coordinate so
that it becomes a simultaneous bang function, i.e., ${\hat t}=0$
at big bang, and hence $t\rightarrow -\infty$ towards the
singularity. We thus take the synchronous coordinates used by
Belinski\v{\i}, Khalatnikov, and Lifshitz (BKL) \cite{bkl82} as
the starting point (see also \cite{walyip81} for a discussion
about the existence of such coordinates) and obtain a
dimensionless formulation that brings us particularly close to the
work of BKL, which therefore can be interpreted directly in terms
of the dimensionless state space picture the present formulation
gives rise to.

It may seem that the latest approach is superior to the
Hubble-normalized one, however, both have advantages and
disadvantages. The advantage of the Hubble-normalized approach is
that one essentially uses the expansion which appears prominently
in the singularity theorems, and that one decouples the
dimensional variable $H$. This implies that $H$ carries the
dimensional constant of integration, which we denote as the scale
parameter even though it is a function in general, when one has a
scale-invariant source. The advantages of the conformal proper
time normalization approach is that it yields a first order system
which may be extended to a first order symmetric hyperbolic
system, and that one obtains a formulation closely related to that
of BKL, which facilitates comparisons. A disadvantage is that one
does not decouple a variable that carries the scale parameter. The
difference of the two approaches as regards the last aspect can be
illustrated by the Kasner subset.

Let us for simplicity only consider the vacuum case (as discussed
in UEWE, if one has a source one may have additional test fields
such as the 3-velocity of a fluid). In the Hubble-normalized
approach the Kasner subset is defined by setting all variables to
zero except the shear which satisfies
$1-\frac{1}{6}\Sig_{\alpha\beta}\,\Sig^{\alpha\beta}=0$ (and $q=2$
if we consider the $r_a$ extension), which yields the so-called
Kasner sphere, see UEWE. On the other hand, when we use the
conformal proper time normalization, with a simultaneous bang
function, then ${\cal H}$ and $\Sig_{\alpha\beta} \neq 0$. Setting
$r_a=(1,0,0,0)$ and all other variables to zero, apart from $\cal
H$ and $\Sig_{\alpha\beta}$, leads to that Eq.~(\ref{dlgauss})
yields: $(1+{\cal
H})^2-\frac{1}{6}\Sig_{\alpha\beta}\,\Sig^{\alpha\beta}=0$, i.e.,
we obtain a cone with ${\cal H}=-1$ as apex. If we consider
expanding models, then ${\cal H}>0$; for Kasner ${\cal H} = {\cal
H}_0 = const>0$, and hence we obtain a Kasner sphere for each
value of ${\cal H}_0$. This illustrates that in contrast to the
Hubble-normalized formulation, we obtain a scale parameter as a
constant of integration in the proper time normalized formulation,
something that somewhat complicates the description of the
structure of the attractor for generic singularities.

Implicit in the above discussion is also the need for choosing a
`dominant' quantity as conformal factor in order to obtain
asymptotically regular and well behaved equations, e.g., if we had
used the inverse of a component of $n^{\alpha\beta}$ as the
conformal factor many state space variables would have blown up
towards an approach to Kasner (and towards a generic singularity).
Thus, e.g., for a generic singularity one needs to use a conformal
factor that goes to zero at least as fast as $H^{-1}$ in order for
the state space variables to remain finite, however, it is
preferable if $\Psi\,H$ remains finite towards the singularity,
i.e., it is preferable to have a `marginally dominant' conformal
factor that leads to finite state space variables, without all of
them going to zero, and well behaved field equations.

\section{Discussion}
\label{discandout}

In this paper we have used conformal transformations in order to
obtain dimensionless regularized field equations that allow one to
extract asymptotic features and properties about the solution
space of general relativity. The conformal factor is to be chosen
so that it captures a characteristic scale associated with
asymptotic structure so that all the state space variables form
dimensionless ratios with respect to this scale. In this paper we
have used a `minimal' approach, however, we could have extended
our formalism to also include the curvature, in particular the
Weyl curvature, and the Bianchi identities, as done in e.g.
\cite{hveugg97} and \cite{ellels99}. Since the Weyl curvature is
conformally invariant, this implies that one should use the Weyl
curvature in a conformal ONF, i.e., in contrast to Friedrich's
conformal approach, see \cite{fri98} and references therein, the
Weyl tensor is not to be scaled with the conformal factor in our
approach---it suffices to express it in a conformal ONF. We here
note that even though the conformal factor enters our equations
implicitly in the combination $\vece_a\Psi/\Psi=r_a$ (Sec. 2
notation), the components of $r_a$ stay {\em finite\/} when
$\Psi\rightarrow 0$, if $\Psi$ is chosen as an appropriate
marginally dominant scalar that carries dimension $q=1$. This
leads to a coupled system of regular dimensionless field
equations, since the equations (differential or algebraically
trivial, depending on the choice of $\Psi$) for the dimensional
$\Psi$ decouple; furthermore, it is the reduced dimensionless
system that carries the essential dynamics, since $\Psi$ can be
obtained afterwards once the equations of the reduced system have
been solved.

Note that the components of the Weyl tensor in an ONF (which have
dimension $q=-2$) and the components of the Weyl tensor in a
conformal ONF (dimension $q=0$ when the conformal factor have
dimension $q=1)$ only differ by the square of the conformal
factor. In \cite{hveugg97} and \cite{book:waiell97} the Weyl
tensor was normalized with $(\sqrt{3}H)^{-2}$ as the scale factor,
since the same factor was used to normalize ${\hat T}^{ab}$, as
discussed previously. However, we now see that from a conformal
point of view the natural normalization factor is just the square
of the conformal factor, which in the Hubble-normalization case is
$H^{-2}$.

Our choice of conformal factor is also quite different than that
used in conformal approaches to isotropic singularities, see e.g.
\cite{goowai85}, \cite{angtod99}. There the motivation for the
conformal factor is a purely mathematical one; choose a conformal
factor so that regular expressions for the covariant coordinate
components of the 3-metric at the singularity are obtained. This
typically leads to a conformal factor that does not have any
particular dimension, indeed, the dimensional weight is different
for different matter sources (not surprisingly, increasingly
complicated dimensional conformal properties lead to increasingly
messy subsequent mathematical analysis---this is why, e.g., the
dimensionally simple case of a perfect fluid with radiation as
equation of state is relatively easy to treat). The present
approach uses a strategy that is almost the opposite. The
conformal factor is always chosen to carry the dimensional weight
and for isotropic singularities, see \cite{limetal04}, as well as
for typical timelines for generic singularities, all the
components of the covariant 3-metric blow up, and this is a very
good thing! Instead the focus is on the components of the spatial
frame vectors which determine the {\em contravariant\/} components
of the 3-metric; these components all go to zero, and this
directly reflects the asymptotic causal properties towards the
singularity---asymptotic silence, see UEWE and \cite{andetal05}.
The present conformal approach emphasizes the geometrical content
of the discussion about asymptotic silence in UEWE and
\cite{andetal05} even further due to the connection between causal
and conformal properties. In the present approach the focus is on
the conformal state space which is extended to include the
so-called silent boundary where all components of the
contravariant 3-metric are zero. This extension then allows one to
use the state space picture to perturb the structure on the silent
boundary into the physical state space, and thus derive physical
results about asymptotic spacetime properties.

In this paper we have used temporal asymptotic aspects associated
with singularities as an example and made contact with other work
to illustrate the usefulness of our approach. However, it is our
belief that the current formalism may be useful for all types of
asymptotics in general relativity: temporal, null, and spacelike;
for non-isolated and isolated systems. Indeed, we already know
that it is useful for future temporal asymptotes in SH contexts,
since it contains the Hubble-normalized formalism which already
have proven to be useful in this regard. However, it should be
pointed out that the Hubble-normalization did not lead directly to
regularized equations towards the future for the general SH
models, some additional manipulations were needed in order to
obtain asymptotic results, but the Hubble-normalization provided
the first key step \cite{waietal99}. A similar situation is
expected for null infinity. The present formalism is expected to
yield direct results for special cases, but not the most general
ones where additional manipulations will be necessary. The current
formalism could have been used as the starting point in the work
\cite{heietal03}, using the radial area coordinate $r$ as
conformal factor, which yielded asymptotic results as regards
spacelike asymptotes, for non-isolated and isolated systems, in
the context of static spherically symmetric spacetimes (in
\cite{heietal03} there existed two relevant scales, however,
several scales can be handled by first using the conformal
transformation to take care of an overall scale, and then making
additional variable transformations that form further ratios,
which one by one quotient out the other scales).

However, the main reason for believing that our proposed approach
may be a useful ingredient in future studies of asymptotics
perhaps comes from the simplicity and naturalness of the main
ideas---summarized as follows:
\begin{itemize}
\item[(i)] Consider a marginally dominant dimensional scale that
captures some key asymptotic features.
\item[(ii)] Use conformal transformations to geometrically
quotient out and decouple this scale so that all remaining
quantities represent dimensionless ratios with respect to that
scale.
\item[(iii)] Use the obtained reduced dimensionless regularized
field equations on an extended state space (i.e., include
asymptotic limits if they occur on the boundary of the original
dimensionless state space) to derive and describe asymptotic
properties.
\end{itemize}
Thus dynamical systems approaches in general relativity, based on
regularized dimensionless field equations, have found their place
in a familiar conformal geometric setting.

\subsection*{Acknowledgements}
It is a pleasure to thank Lars Andersson, Henk van Elst, Woei Chet
Lim, and John Wainwright for many helpful and stimulating
asymptotically silent discussions. CU is supported by the Swedish
Research Council.

\end{document}